\documentclass[12pt,floatfix]{revtex4}
\usepackage{graphicx}
\usepackage{amsmath}

\begin{document}

\title{Fission of charged alanine dipeptides}

\author{Alexander V. Yakubovitch\dag,
        Ilia A. Solov'yov\dag\ddag, 
        Andrey V. Solov'yov\dag\ddag, 
        and Walter Greiner\ddag}

\address{\dag\ A.F. Ioffe Physical-Technical Institute, Russian
Academy of Sciences, Politechnicheskaja str. 26, 194021 St Petersburg,  Russia}  

\address{\ddag\ Institut f\"{u}r Theoretische Physik der Universit\"{a}t
Frankfurt am Main, Robert-Mayer Str. 8-10, D-60054 Frankfurt am Main, Germany}

\email{solovyov@th.physik.uni-frankfurt.de,ilia@th.physik.uni-frankfurt.de}

\begin{abstract}
In this work we have performed for the first time a systematic analysis
of the dissociation and fission pathways of neutral, singly and doubly charged
alanine dipeptide ions with the aim to identify the fission mechanism and the most
probable fragmentation channels of these type of molecules.
We demonstrate the importance of rearrangement of the molecule structure during the fission process.
This rearrangement may include transition to another isomer or a quasi-molecular state before
actual separation of the daughter fragments begins.
\end{abstract}

\maketitle

\section{Introduction}

{\it This work was presented on "The eighth European Conference on Atomic and Molecular Physics"
(ECAMPVIII) (Rennes, France, July 6-10, 2004) and on the 
"Electronic Structure Simulations of Nanostructures" workshop (ESSN2004)
(Jyv\"askyl\"a Finland, June 18-21, 2004).}
\\
\\

Amino acids are building blocks for proteins. Recently, it became possible to study experimentally
fragments of proteins, i.e. chains of amino acids, in a gas phase with the use of the MALDI
mass spectroscopy \cite{Beavis96,Cohen96,Karas88}.
{\it Ab initio} theoretical investigations of amino acid chains began also only recently
\cite{Salahub01,Pliego03,Srinivasan02,Head-Gordon91,Gould94,Beachy97,LesHouches}
and are still in its infancy.
In this work we have performed for the first time a systematic analysis
of the dissociation and fission pathways of neutral, singly and doubly charged
alanine dipeptide ions with the aim to identify the fission mechanism and the most
probable fragmentation channels of these type of molecules.
We demonstrate the importance of rearrangement of the molecule structure during the fission process.
This rearrangement may include transition to another isomer or a quasi-molecular state before
actual separation of the daughter fragments begins.

\section{Theoretical methods}

Our exploration of the energetics and the pathways of the
fragmentation process is based on the density-functional theory (DFT)
and the post-Hartree-Fock many-body perturbation theory accounting for all electrons in
the system. Within the DFT one has to solve the Kohn-Sham equations, which read as
(see e.g. \cite{MetCl99,LesHouches}):

\begin{equation}
\left( \frac{\hat p^2}2+U_{ions}+V_{H}+V_{xc}\right)
\psi_i =\varepsilon _i \psi _i,
\end{equation}
where the first term represents the kinetic energy of the $i$-th electron,
and $U_{ions}$ describes its attraction to the ions in the cluster,
$V_{H}$ is the Hartree part of the interelectronic interaction:
\begin{equation}
V_{H}(\vec r)=\left. \int \frac{\rho(\vec r\,')}{|\vec r-\vec r\,'|}
\, d\vec r\,'\right.,
\end{equation}
and $\rho(\vec r\,')$ is the  electron density:
\begin{equation}
\rho(\vec r)=\sum_{\nu=1}^{N} \left|\psi_i(\vec r) \right|^2,
\end{equation}

\noindent
where $V_{xc}$ is the local exchange-correlation potential,
$\psi_i$ are the electronic orbitals and $N$ is the number of 
electrons in the cluster.

The exchange-correlation potential is defined as the functional
derivative of the exchange-correlation energy functional:
\begin{equation}
V_{xc}=\frac{\delta E_{xc}[\rho]}{\delta \rho(\vec r)},
\end{equation}

The approximate functionals employed by DFT methods partition the
exchange-correlation energy into two parts, referred to as exchange
and correlation parts. Both parts are the functionals of
the electron density, which can be of two distinct types: either local functional
depending on only the electron density $\rho$ or gradient-corrected functionals depending on both
$\rho$ and its gradient, $\nabla\rho$. In literature, there is a variety of exchange correlation
functionals. In our work we use the Becke's three parameter gradient-corrected exchange
functional with the gradient-corrected correlation functional of Lee, Yang and Parr (B3LYP)
\cite{Becke88,LYP,Parr-book}. 
We utilize the standard 6-31G(d) basis set to expand the electronic orbitals $\psi_i$.

To explore the energetic and the pathways of the fragmentation process we introduced the procedure
which implies the calculation of the multidimensional potential energy surface. We increase 
the distance between the two fragments and calculate the multidimensional potential energy surface
on each step. Then we identify local minima on this surface for a given distance between the fragments,
and so find the fragmentation pathways of the molecule. For our calculations we utilized 
the Gaussian 03 software package \cite{Gaussian03,Gaussian98_man}.

\section{Results of calculation}
\begin{figure}[h]
\includegraphics[scale=0.5]{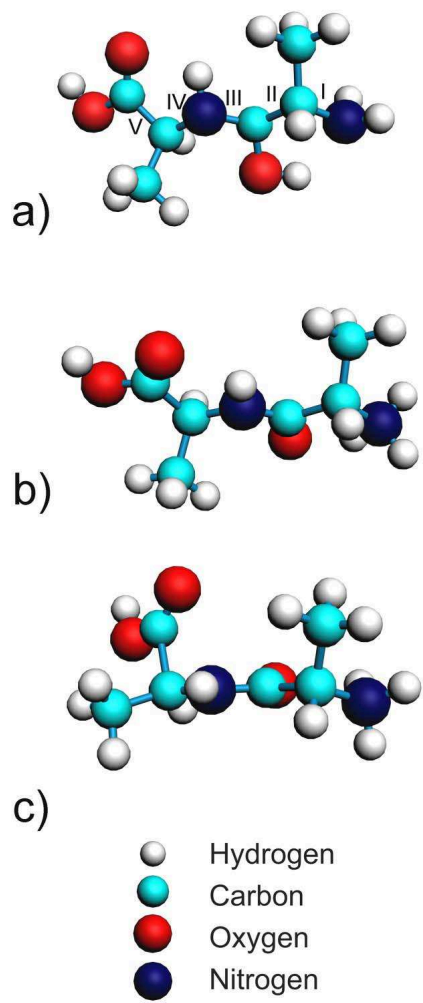}
\caption{Optimized geometries of neutral (part a), singly charged (part b)
and doubly charged (part c) alanine dipeptides terminated  with the
$-NH_3$ radical.}
\label{geom_NH3}
\end{figure}

\begin{figure}[h]
\includegraphics[scale=0.4]{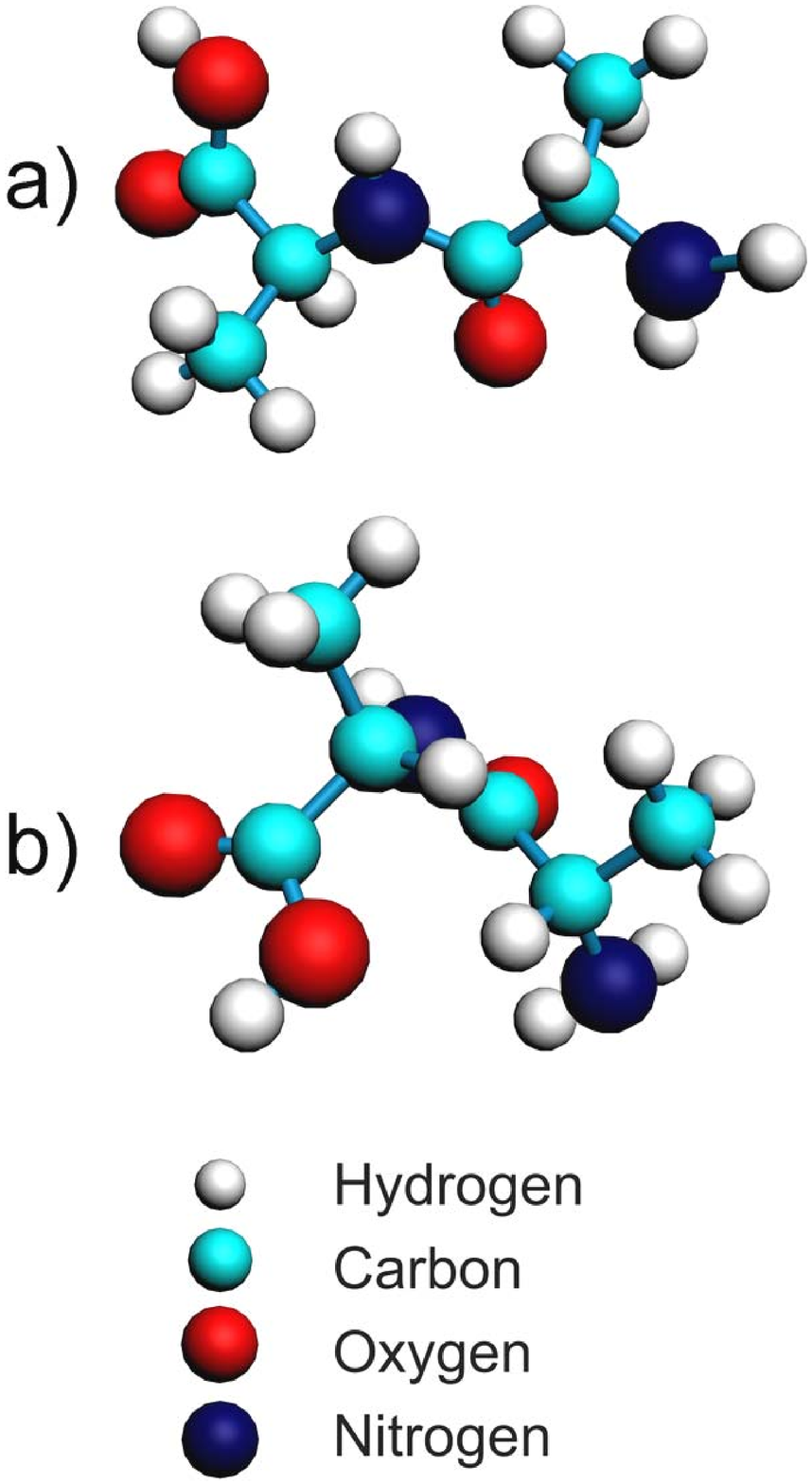}
\caption{Optimized geometries of neutral (part a) and singly charged (part b)
alanine dipeptides terminated  with the $-NH_2$ radical.}
\label{geom_NH2}
\end{figure}

In our work we have investigated the fission process of the two types of the alanine dipeptide.
The only difference between them is in the $N-$ ending of the molecule, which could be terminated
by the $-NH_2$ or $-NH_3$ radicals. The geometries of the optimized neutral and charged
alanine dipeptides terminated with the
$-NH_3$ and $-NH_2$ radicals are shown in figures \ref{geom_NH3} and \ref{geom_NH2}
respectively. In figure \ref{geom_NH3} we number different fragmentation
channels along the polypeptide chain by Roman numerals. These fragmentation channels is
a subject for investigation in our work.

\begin{figure}[h]
\includegraphics[scale=0.71]{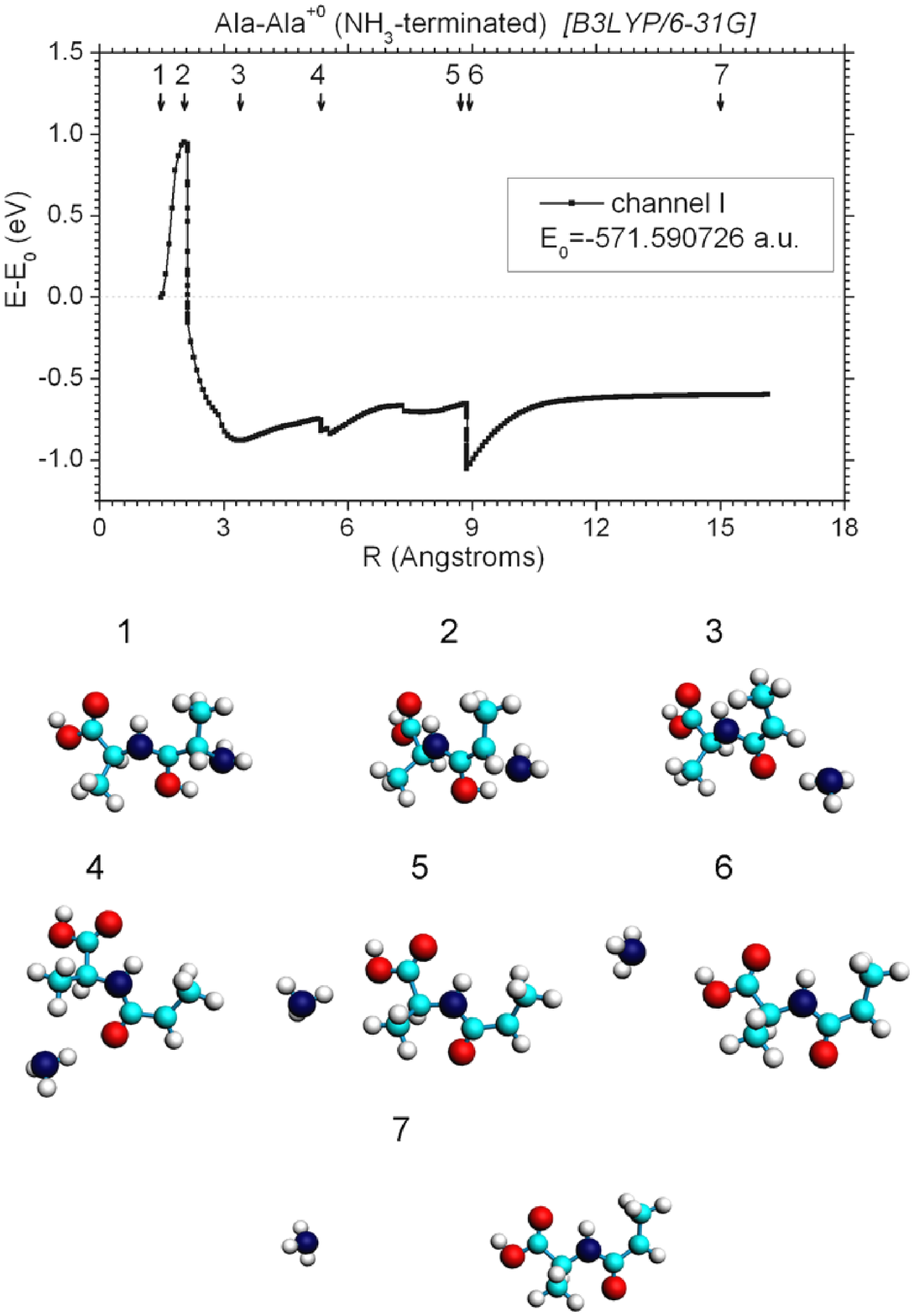}
\caption{Total energy of neutral alanine dipeptide terminated by the $-NH_3$
radical as a function of distance between two fragments for fragmentation channel I
(see figure \ref{geom_NH3}) calculated with the B3LYP method. The energy scale is given
in eV, counted from the ground state energy of the optimized chain configuration
of the neutral alanine dipeptide,
$E_0^{NH_3}=-571.590726$ a.u.
The numbered geometries correspond to the marked points
on the energy curve.}
\label{fig_0_I}
\end{figure}

\begin{figure}[h]
\includegraphics[scale=0.71]{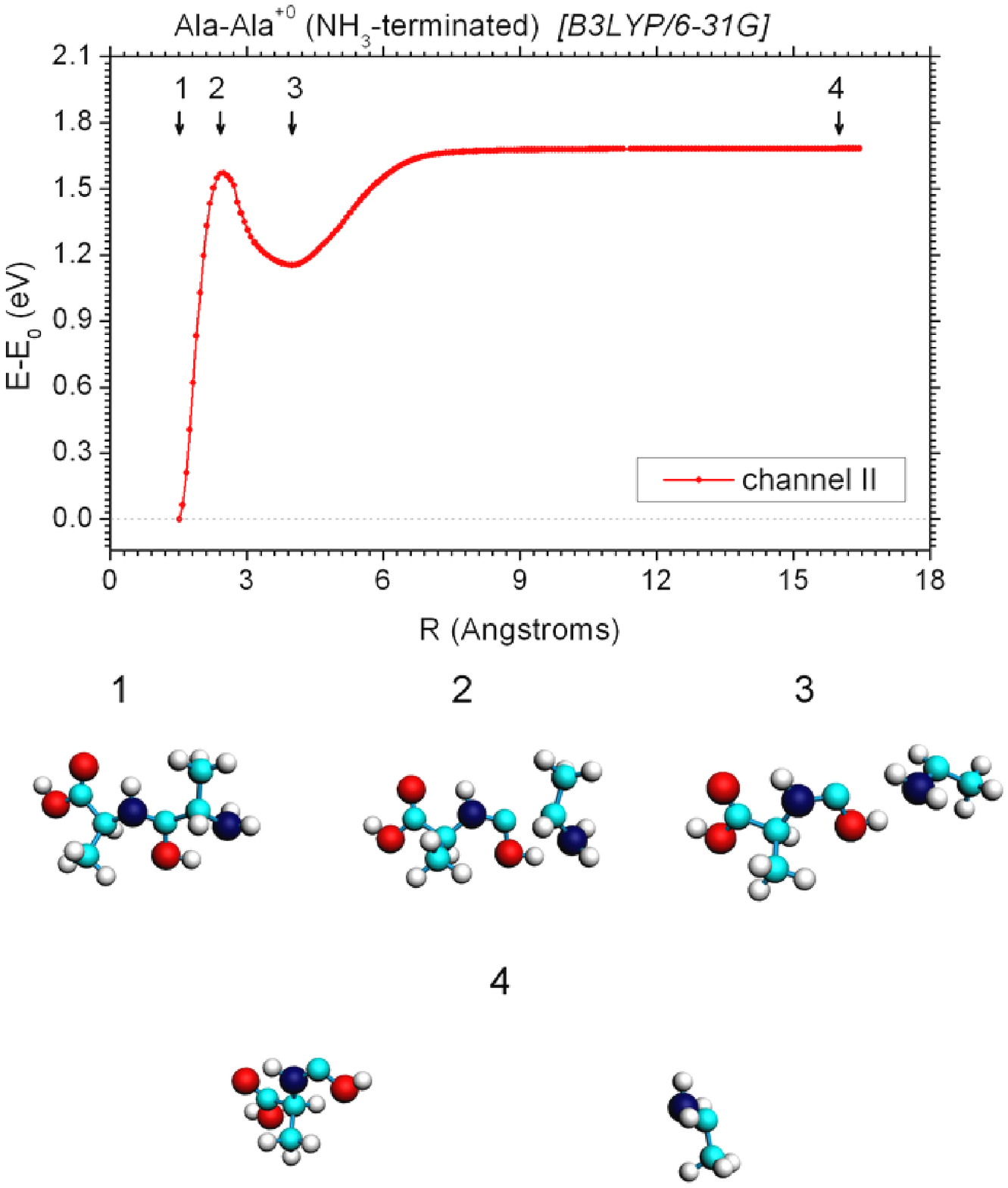}
\caption{Total energy of neutral alanine dipeptide terminated by the $-NH_3$
radical as a function of distance between two fragments for fragmentation channel II
(see figure \ref{geom_NH3}) calculated with the B3LYP method.
The energy scale is given
in eV, counted from the ground state energy of the optimized chain configuration
of the neutral alanine dipeptide,
$E_0^{NH_3}=-571.590726$ a.u.
The numbered geometries correspond to the marked points
on the energy curve.}
\label{fig_0_II}
\end{figure}

\begin{figure}[h]
\includegraphics[scale=0.71]{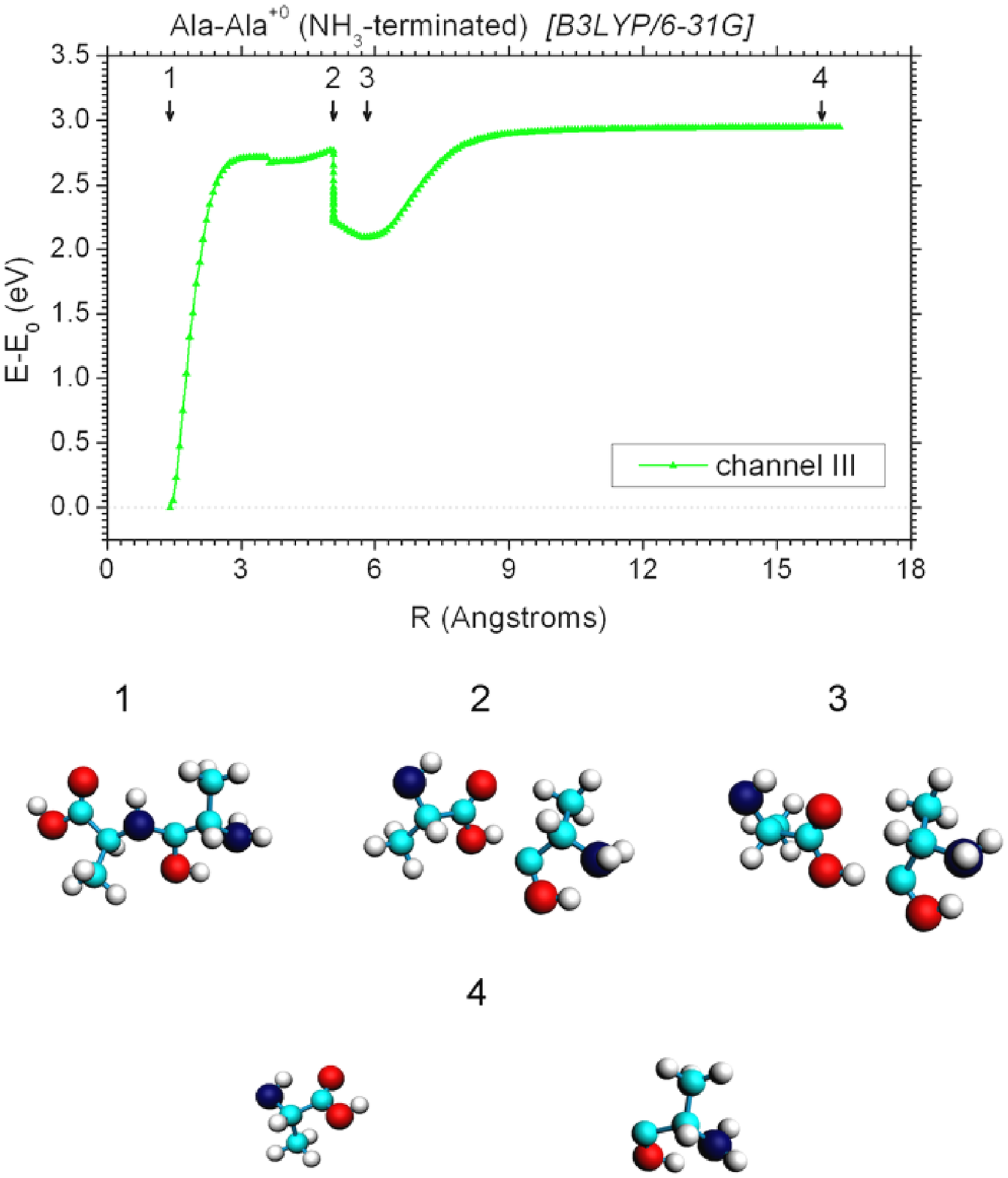}
\caption{Total energy of neutral alanine dipeptide terminated by the $-NH_3$
radical as a function of distance between two fragments for fragmentation channel III
(see figure \ref{geom_NH3}) calculated with the B3LYP method. The energy scale is given
in eV, counted from the ground state energy of the optimized chain configuration
of the neutral alanine dipeptide,
$E_0^{NH_3}=-571.590726$ a.u.
The numbered geometries correspond to the marked points
on the energy curve.}
\label{fig_0_III}
\end{figure}

\begin{figure}[h]
\includegraphics[scale=0.71]{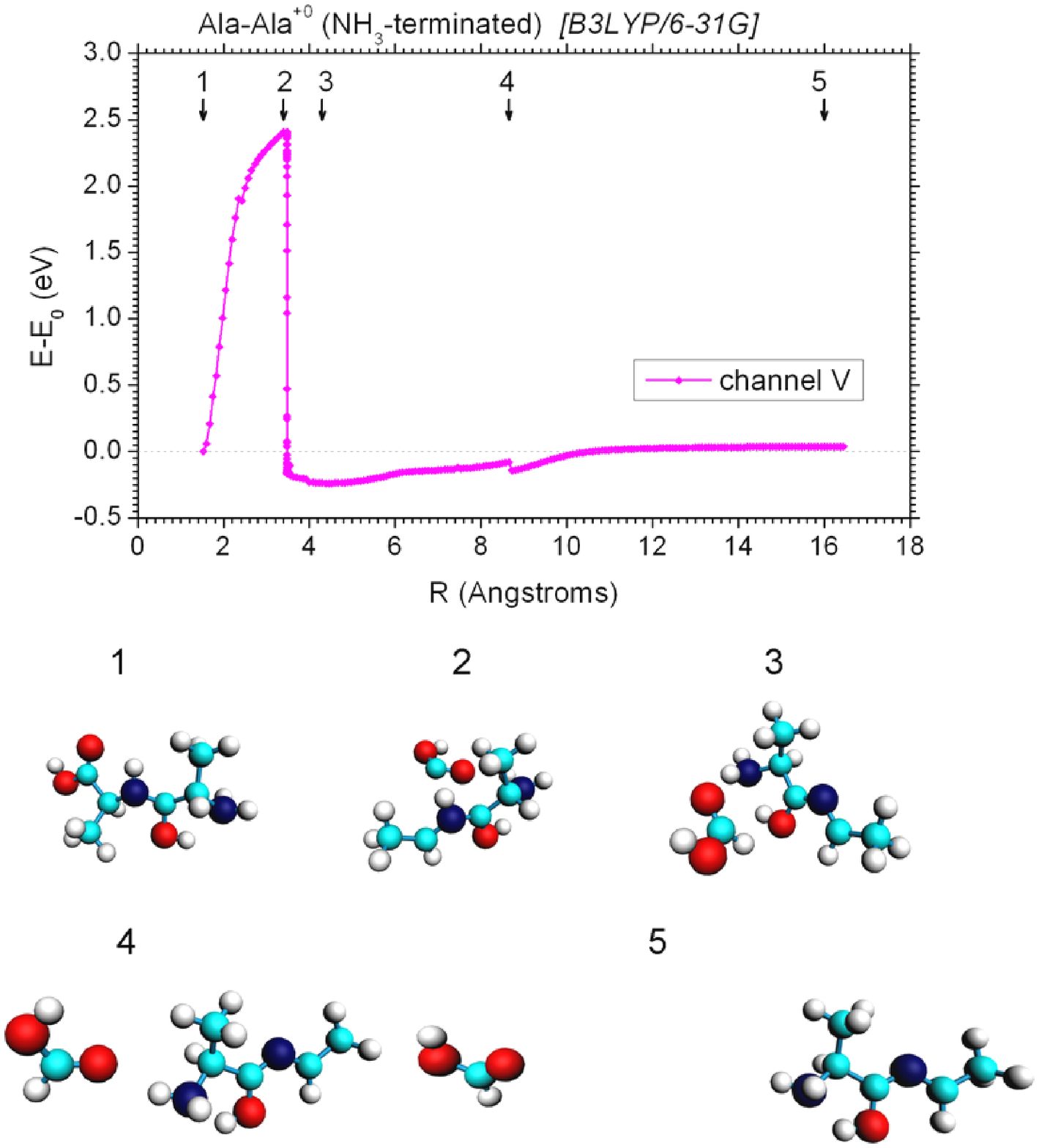}
\caption{Total energy of neutral alanine dipeptide terminated by the $-NH_3$
radical as a function of distance between two fragments for fragmentation channel V
(see figure \ref{geom_NH3}) calculated with the B3LYP method.
The energy scale is given
in eV, counted from the ground state energy of the optimized chain configuration
of the neutral alanine dipeptide,
$E_0^{NH_3}=-571.590726$ a.u.
The numbered geometries correspond to the marked points
on the energy curve.}
\label{fig_0_V}
\end{figure}

\begin{figure}[h]
\includegraphics[scale=0.71]{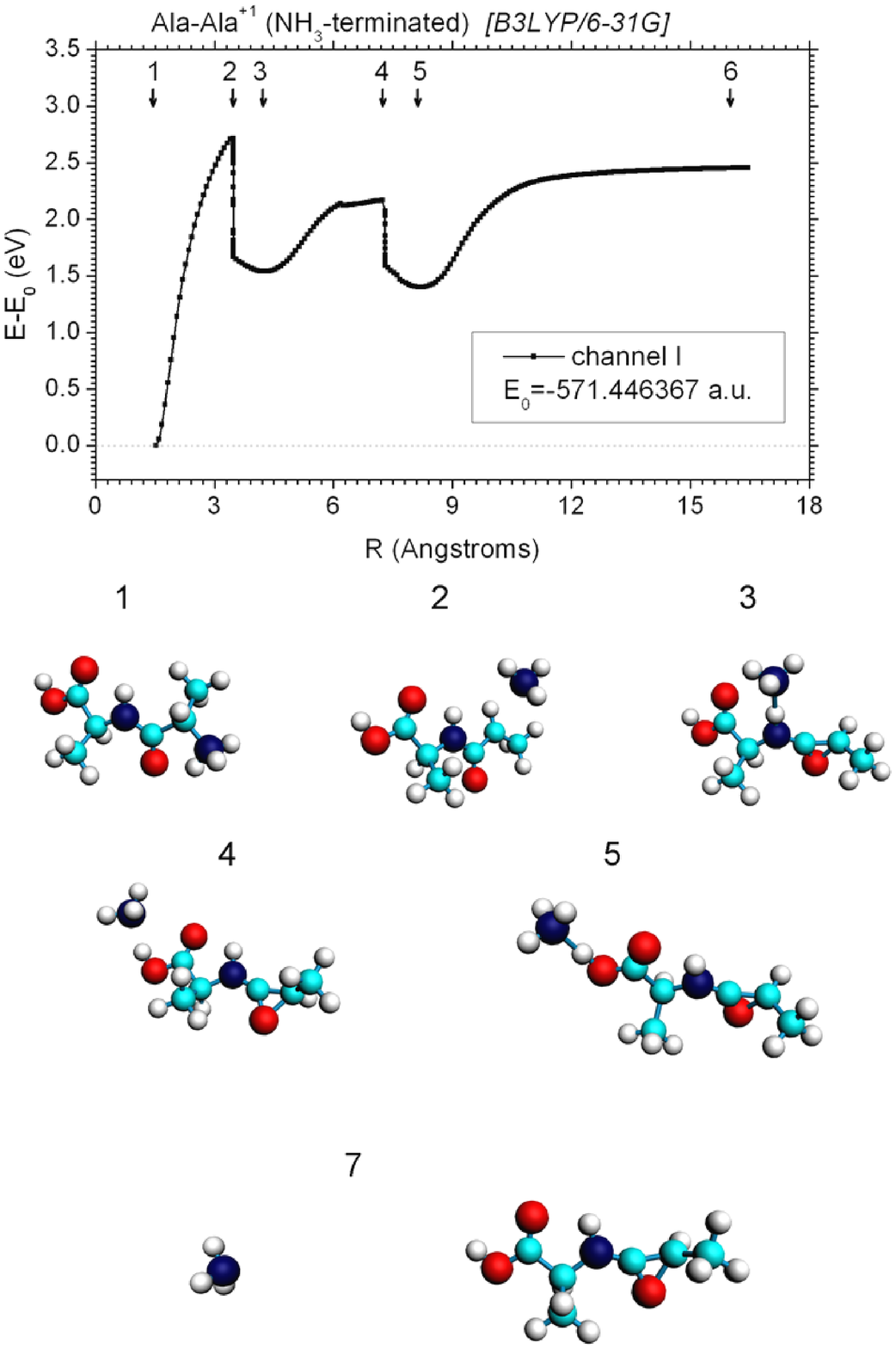}
\caption{Total energy of singly charged alanine dipeptide terminated by the $-NH_3$
radical as a function of distance between two fragments for fragmentation channel I
(see figure \ref{geom_NH3}) calculated with the B3LYP method.
The energy scale is given
in eV, counted from the ground state energy of the optimized chain configuration
of the singly charged alanine dipeptide,
$E_{+1}^{NH_3}=-571.446367$ a.u.
The numbered geometries correspond to the marked points
on the energy curve.}
\label{fig_1_I}
\end{figure}

\begin{figure}[h]
\includegraphics[scale=0.71]{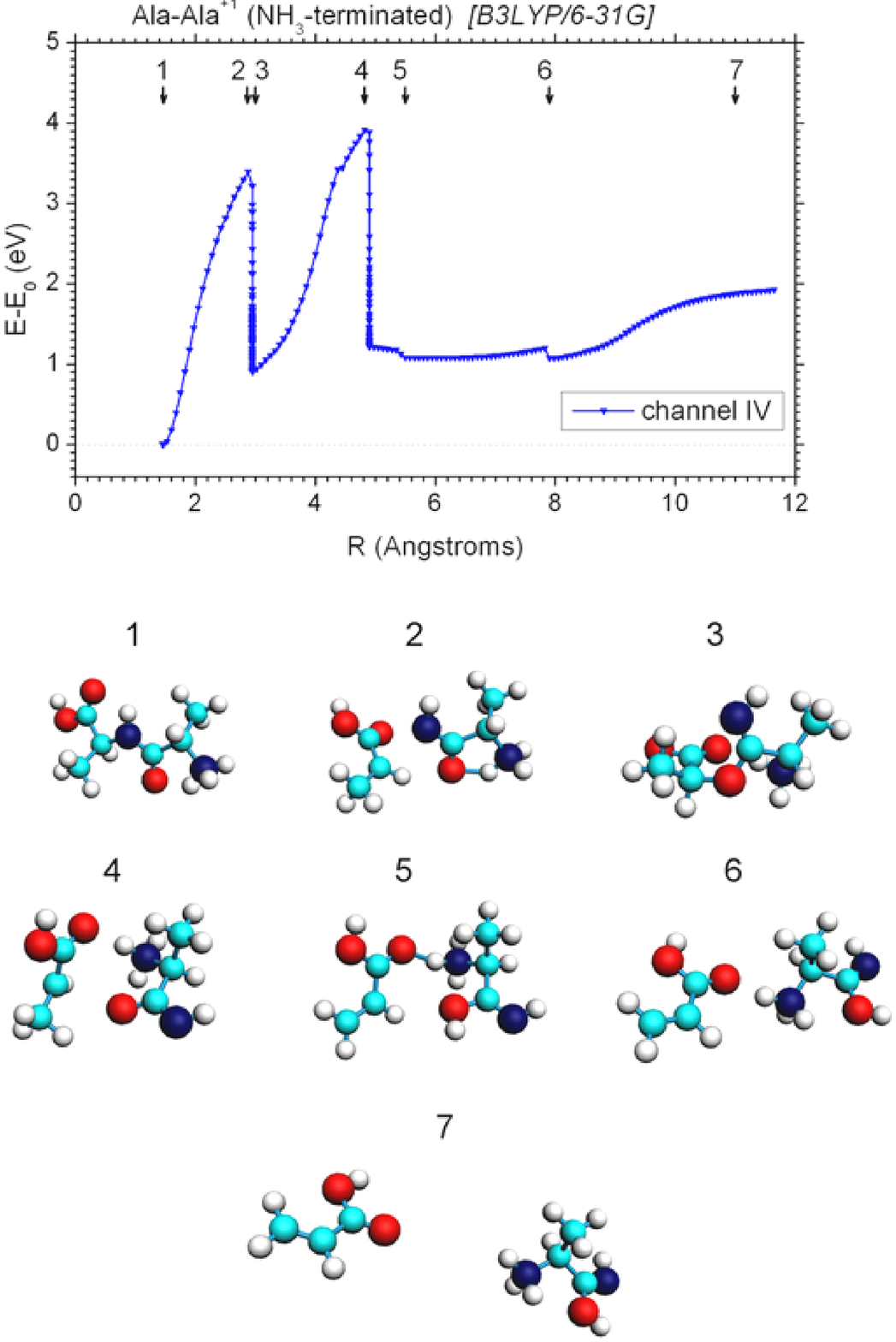}
\caption{Total energy of singly charged alanine dipeptide terminated by the $-NH_3$
radical as a function of distance between two fragments for fragmentation channel IV
(see figure \ref{geom_NH3}) calculated with the B3LYP method.
The energy scale is given
in eV, counted from the ground state energy of the optimized chain configuration
of the singly charged alanine dipeptide,
$E_{+1}^{NH_3}=-571.446367$ a.u.
The numbered geometries correspond to the marked points
on the energy curve.}
\label{fig_1_IV}
\end{figure}

\begin{figure}[h]
\includegraphics[scale=0.71]{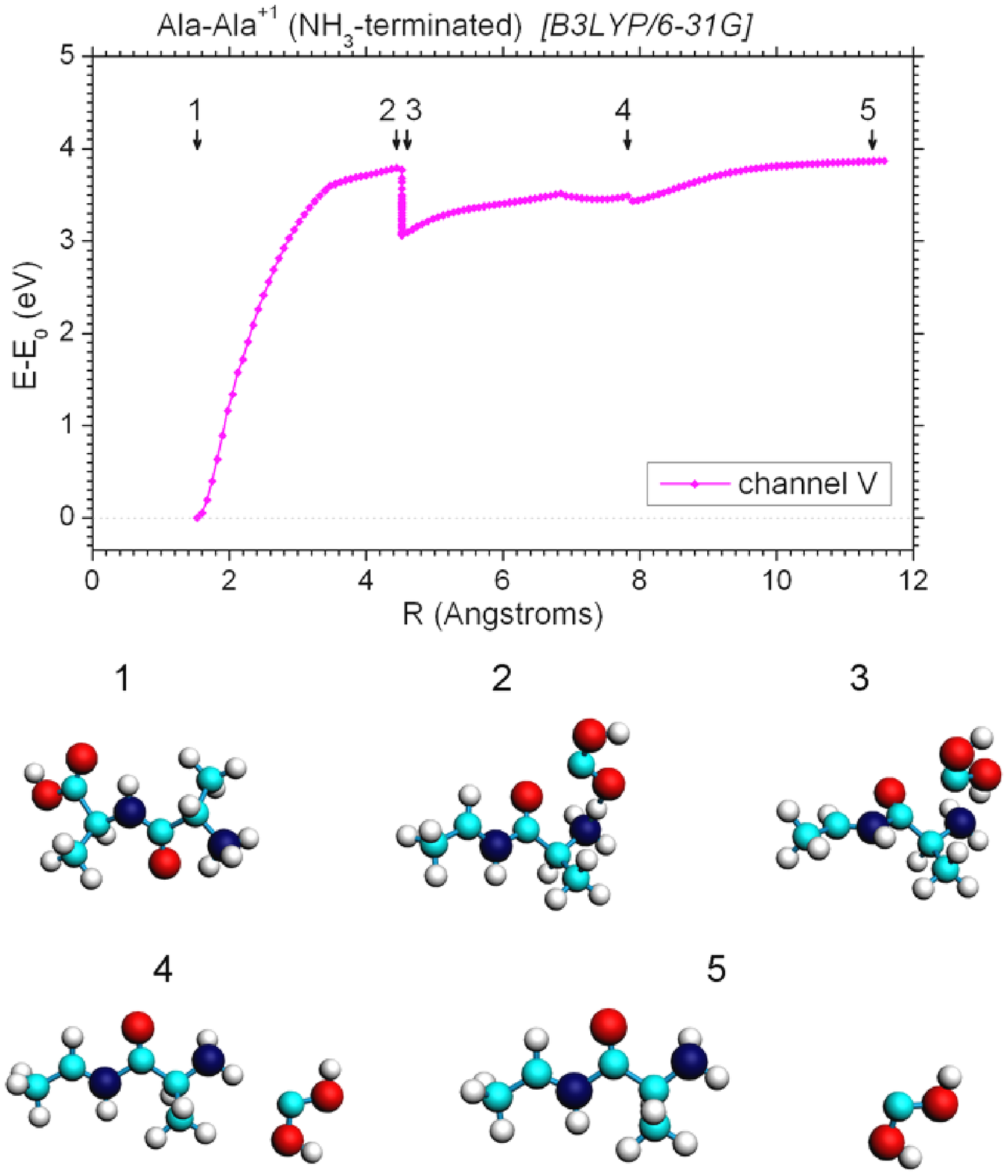}
\caption{Total energy of singly charged alanine dipeptide terminated by the $-NH_3$
radical as a function of distance between two fragments for fragmentation channel V
(see figure \ref{geom_NH3}) calculated with the B3LYP method.
The energy scale is given
in eV, counted from the ground state energy of the optimized chain configuration
of the singly charged alanine dipeptide,
$E_{+1}^{NH_3}=-571.446367$ a.u.
The numbered geometries correspond to the marked points
on the energy curve.}
\label{fig_1_V}
\end{figure}

\begin{figure}[h]
\includegraphics[scale=0.71]{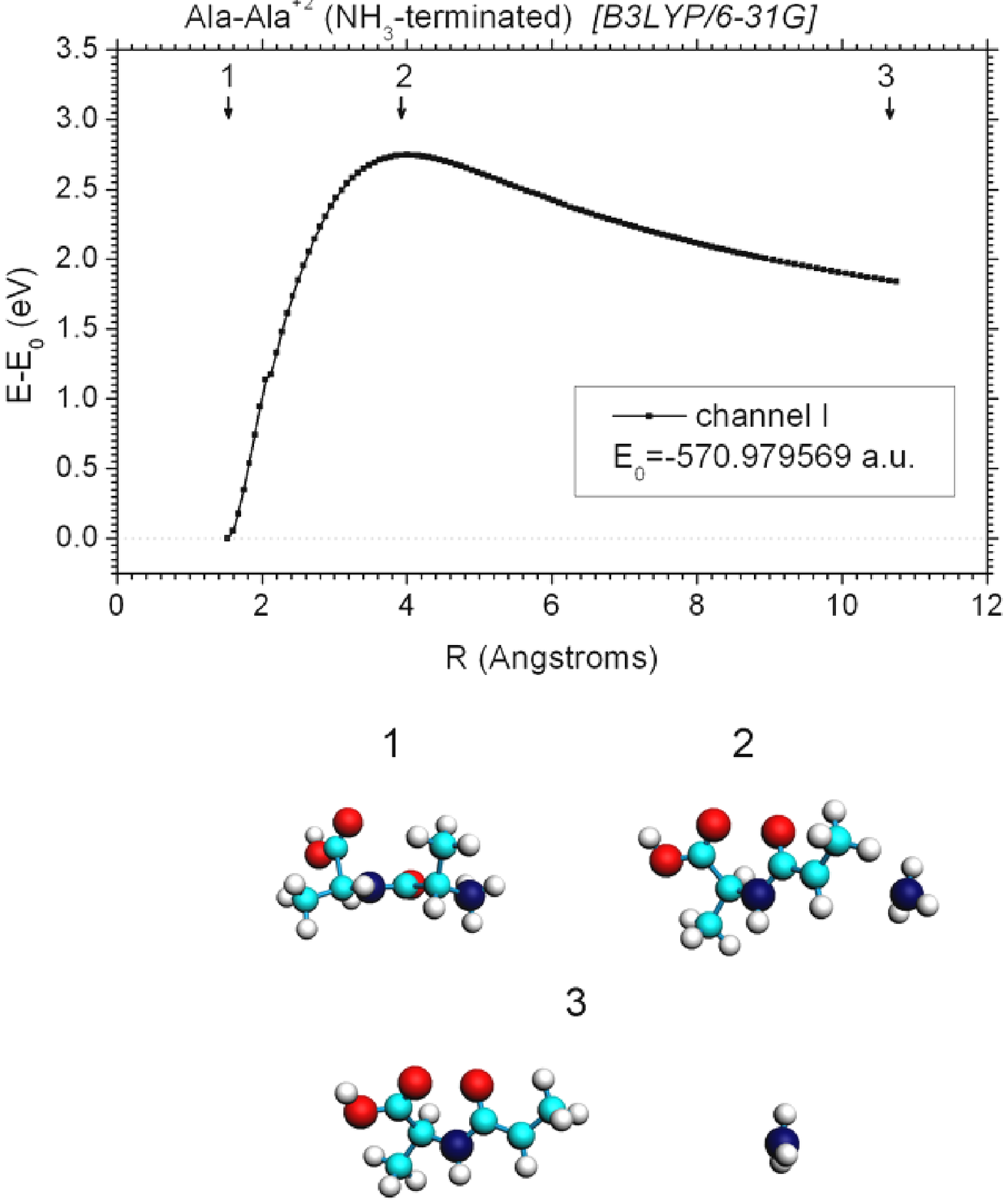}
\caption{Total energy of doubly charged alanine dipeptide terminated by the $-NH_3$
radical as a function of distance between two fragments for fragmentation channel I
(see figure \ref{geom_NH3}) calculated with the B3LYP method.
The energy scale is given
in eV, counted from the ground state energy of the optimized chain configuration
of the doubly charged alanine dipeptide,
$E_{+2}^{NH_3}=-570.979569$ a.u.
The numbered geometries correspond to the marked points
on the energy curve.}
\label{fig_2_I}
\end{figure}

\begin{figure}[h]
\includegraphics[scale=0.71]{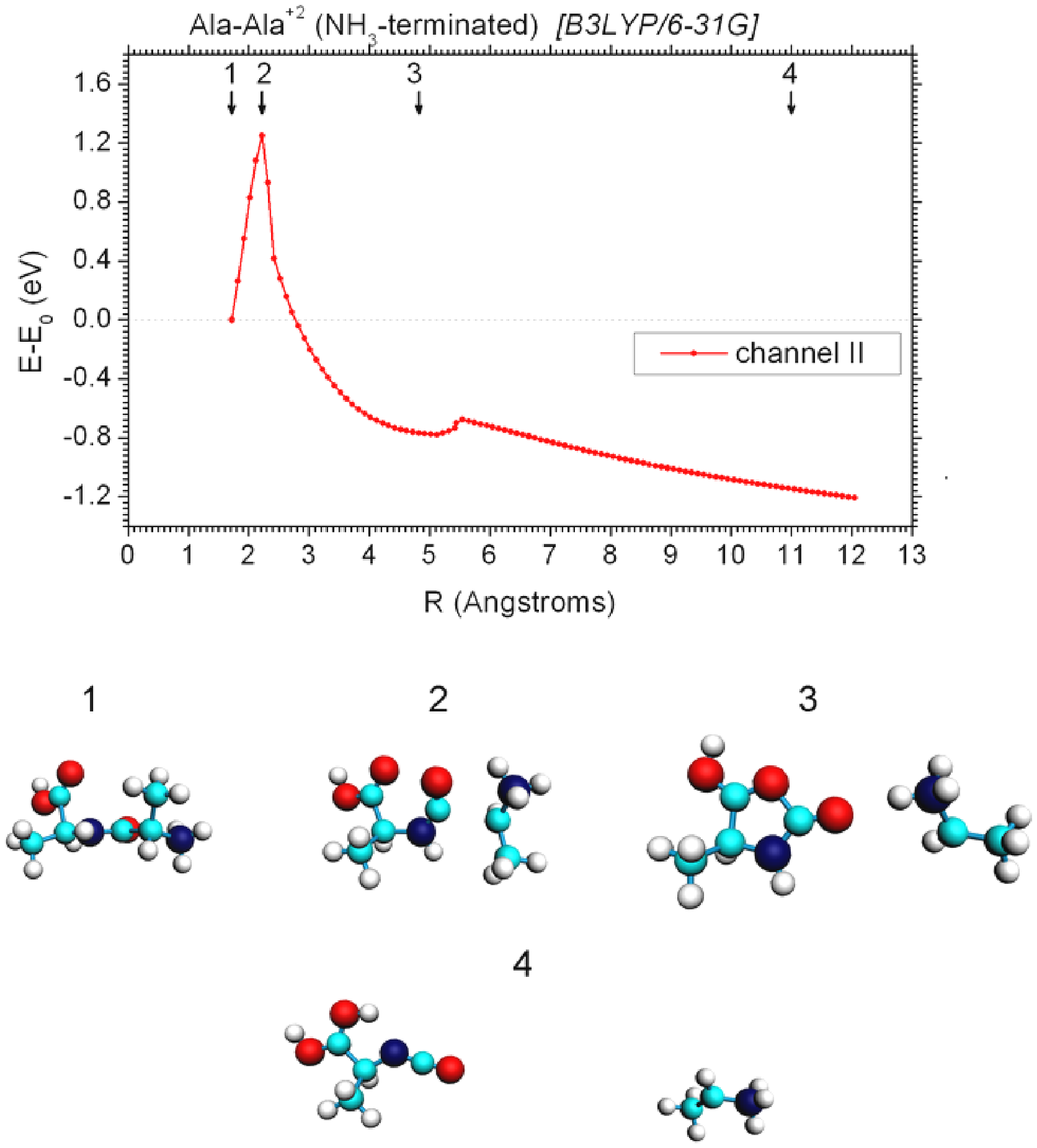}
\caption{Total energy of doubly charged alanine dipeptide terminated by the $-NH_3$
radical as a function of distance between two fragments for fragmentation channel II
(see figure \ref{geom_NH3}) calculated with the B3LYP method.
The energy scale is given
in eV, counted from the ground state energy of the optimized chain configuration
of the doubly charged alanine dipeptide,
$E_{+2}^{NH_3}=-570.979569$ a.u.
The numbered geometries correspond to the marked points
on the energy curve.}
\label{fig_2_II}
\end{figure}

The dependence of the total energy of neutral and charged alanine dipeptide
terminated by the $-NH_3$ radical as a function of distance $R$ between two fragments'
for various fragmentation channels is shown in the figures \ref{fig_0_I}-\ref{fig_2_II}.

Our calculations show that the fragmentation of the alanine dipeptide is a nonmonotoneous
process. The chain configuration of the neutral dipeptide appears to be not the ground state
of the molecule as it is clear from the rearangement process occuring in the channels I and V.
The dipeptide fragmentation process for these channels is exothermic. The single ionization of
the chain configuration of the dipeptide increases its stability and the fragmentation process
in this case becomes endothermic. 

The doubly charged alanine dipeptide however becomes unstable because of  the Coulomb repulsive
forces. This fact is illustrated by the energy dependence on $R$ for channel II. From this dependence
it is seen that the total energy of the system in the final state of the fragmentation process
decreases as compared to its initial state, which means that the doubly charged alanine dipeptide
ion terminated with the $-NH_3$ radical is unstable towards fission, which should take place if the
temperature of the molecule is sufficiently high.

Note that for many fragmentation channels the energy dependence exhibits a barrier
calculation of which takes account of the molecular rearrangement. Such rearrangement
may occur before the actual separation of the daughter fragments begins. In many cases the
rearrangement leads to the formation of a quasimolecular state
(see figures  \ref{fig_0_I}-\ref{fig_2_II}). 

\begin{figure}[h]
\includegraphics[scale=0.71]{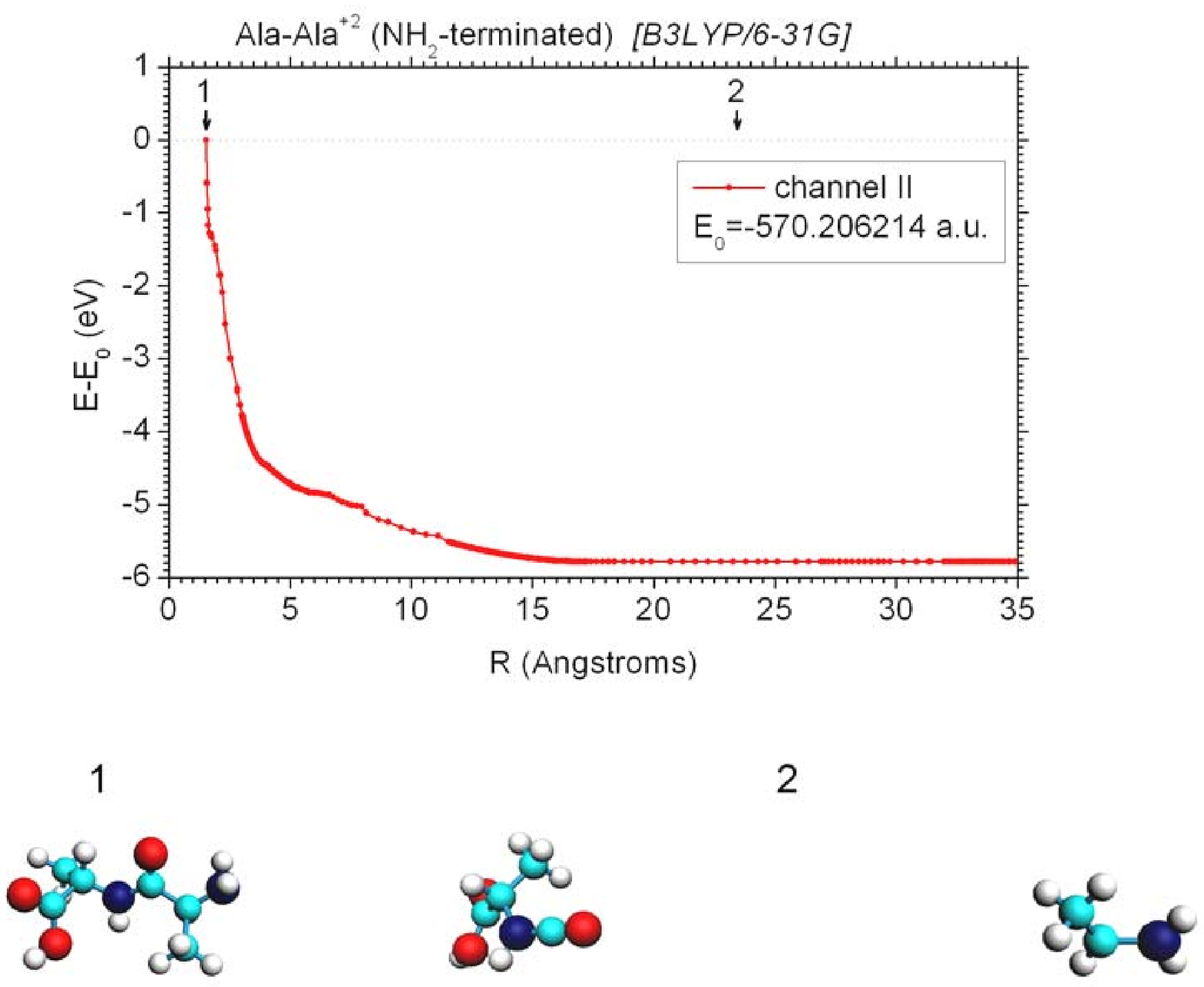}
\caption{Total energy of doubly charged alanine dipeptide terminated by the $-NH_2$
radical as a function of distance between two fragments for fragmentation channel II
(see figure \ref{geom_NH2}) calculated with the B3LYP method. The energy scale is given
in eV, counted from the energy $E_{+2}^{NH_3}=-570.206214$ a.u.
The numbered geometries correspond to the marked points
on the energy curve.}
\label{fig_2_II_NH2}
\end{figure}

\begin{figure}[h]
\includegraphics[scale=0.71]{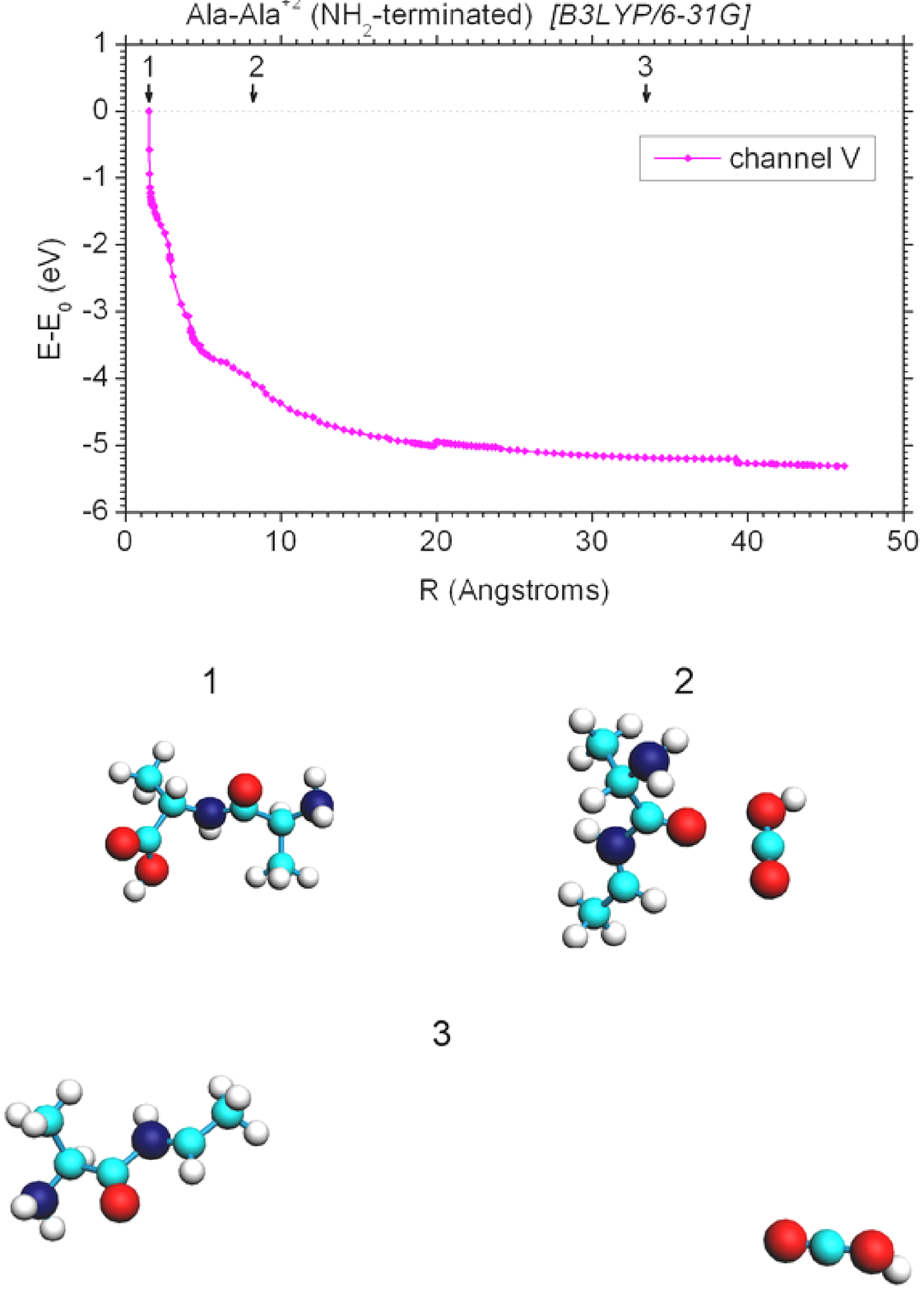}
\caption{Total energy of doubly charged alanine dipeptide terminated by the $-NH_2$
radical as a function of distance between two fragments for fragmentation channel V
(see figure \ref{geom_NH2}) calculated with the B3LYP method. The energy scale is given
in eV, counted from the energy $E_{+2}^{NH_3}=-570.206214$ a.u.
The numbered geometries correspond to the marked points
on the energy curve.}
\label{fig_2_V_NH2}
\end{figure}

Fragmentation of the alanine dipeptide terminated with the $-NH_2$ radical occurs
differently. Thus the fragmentation of the doubly charged ions of this kind
takes place without any fission barrier. In the figures  \ref{fig_2_II_NH2}-\ref{fig_2_V_NH2}
we present the dependence of the dipeptide energy on $R$ for the fission channel II and V in
which the energy release is maximum.

\end{document}